\documentclass[final,1p,times]{elsarticle} 
\usepackage{bm,amsmath,amssymb}
\usepackage{graphicx}
\usepackage{subfigure}

\newcommand \ie {{\sl i.e.}}

\newcommand \etal {{\sl et al.}}

\newcommand \fig[1] {Fig.\ \ref{#1}}

\newcommand \eq[1] {Eq.\ (\ref{#1})}
\newcommand \beq {\begin{equation}}
\newcommand \eeq {\end{equation}}
\newcommand \beqa {\begin{eqnarray}}
\newcommand \eeqa {\end{eqnarray}}
\newcommand \bsubeq {\begin{subequations}}
\newcommand \esubeq {\end{subequations}}
\newcommand \benum {\begin{enumerate}}
\newcommand \eenum {\end{enumerate}}
\newcommand \bitem {\begin{itemize}}
\newcommand \eitem {\end{itemize}}
\newcommand \bfig {\begin{figure}[!t]\begin{center}}
\newcommand \efig {\end{center}\end{figure}}
\newcommand \btab {\begin{table}[!t]\begin{center}}
\newcommand \etab {\end{center}\end{table}}

\newcommand \lang {\left\langle}
\newcommand \rang {\right\rangle}

\journal{Nuclear Physics A} 
\begin{document}
\begin{frontmatter} 

\title{Freeze-out Conditions from Lattice QCD}
\author{Swagato Mukherjee}
\address{Physics Department, Brookhaven National Laboratory, Upton, NY 11973, U.S.A.}

\begin{abstract} 
We describe a procedure for determination of freeze-out parameters of heavy-ion
collisions through direct comparisons between experimentally measured higher order
cumulants of charge fluctuations and first principle (lattice) QCD calculations. 
\end{abstract} 

\end{frontmatter}
\section{Introduction}

The QCD critical point is a conjectured second order phase transition point in the
temperature ($T$) and baryon chemical potential ($\mu_B$) plane of the QCD phase
diagram. Being a second order phase transition point it will be associated with large
correlation lengths, which in turn will be manifested in characteristic large
fluctuations. In this vein, one of the major focus in search for the QCD critical
point in the Relativistic Heavy-Ion Collider (RHIC)'s Beam Energy Scan (BES) program
has been concentrated on measuring higher order cumulants of conserved charge
fluctuations, such as net baryon ($B$), net electric charge ($Q$) and net strangeness
($S$). 

Experimentally measured hadronic observables in Heavy-Ion Collisions (HIC)
characterize the freeze-out condition. The success of statistical hadronization
models, based on the thermal hadron spectrum of the Hadrom Resonance Gas (HRG) model,
in fitting the experimentally measured hadron yields suggests that freeze-out
conditions in HIC can be described by equilibrium thermodynamics characterized by
freeze-out temperatures ($T^f$) and chemical potentials ($\mu_B^f,\mu_Q^f,\mu_S^f$).
On the other hand, the QCD critical point will be located along the QCD
transition/crossover line in the $T-\mu_B$ plane. Thus in order to observe signatures
of the critical point in RHIC BES, the freeze-out of the conserved charge
fluctuations must happen at some $(T^f,\mu_B^f)$ close to the QCD transition line in
the $T-\mu_B$ plane. While for moderate values of $\mu_B$ the QCD transition line in
the $T-\mu_B$ plane is known from first principle Lattice QCD (LQCD) calculations
\cite{Bazavov:2011nk,Kaczmarek:2011zz}, till now there is no first principle QCD
determination of freeze-out parameters associated with the observables related to the
conserved charge fluctuations. 

Here we outline a procedure for the determination of the freeze-out parameters
$(T^f,\mu_B^f,\mu_Q^f,\mu_S^f)$ via direct comparison between first principle LQCD
calculations and the experimentally measured cumulants of charge fluctuations. While
such a procedure will tell us whether the experimentally measured fluctuations can
indeed be described by equilibrium thermodynamics, the freeze-out parameters obtained
using this procedure do not necessarily correspond to the chemical freeze-out
parameters. Chemical freeze-out indicate irrelevance of inelastic scatterings, but
the freeze-out of the conserve charge fluctuations takes place through diffusion
processes. It is likely that hadronic final state interactions are important for the
freeze-out of charge fluctuations and it is a-priori not clear that the charge
fluctuations freeze-out at the chemical freeze-out point. Thus, in order to make
ab-initio parameter-free theoretical predictions relevant for the RHIC BES it is
crucial to know the freeze-out parameters associated with the conserved charge
fluctuations.  Once these freeze-out parameters are known from the comparison of the
lower order cumulants only then higher order cumulants can be predicted from LQCD
calculations along this freeze-out line in a completely parameter-free manner.
Subsequent comparisons with the experimental data will clarify to what extent the
higher order cumulants contain non-critical/critical signatures.

The LQCD computations that will be presented here were performed using 2+1 flavor
Highly Improved Staggered Quarks (HISQ) with a physical strange quark mass and light
quark masses corresponding to the Goldstone pion mass of 160 MeV. Three different
lattice spacings corresponding to the temporal extents $N_\tau=6,8,12$ were used for
the calculations. Further details concerning the LQCD calculations as can be found in
Refs. \cite{Bazavov:2011nk,Bazavov:2012jq}. More detailed discussions regarding the
determination of the freeze-out parameters can be found in Ref.
\cite{Bazavov:2012vg}.

\section{Electric charge and strangeness chemical potentials}

\bfig
\subfigure[]{ \label{fig:qi} \includegraphics[scale=0.32]{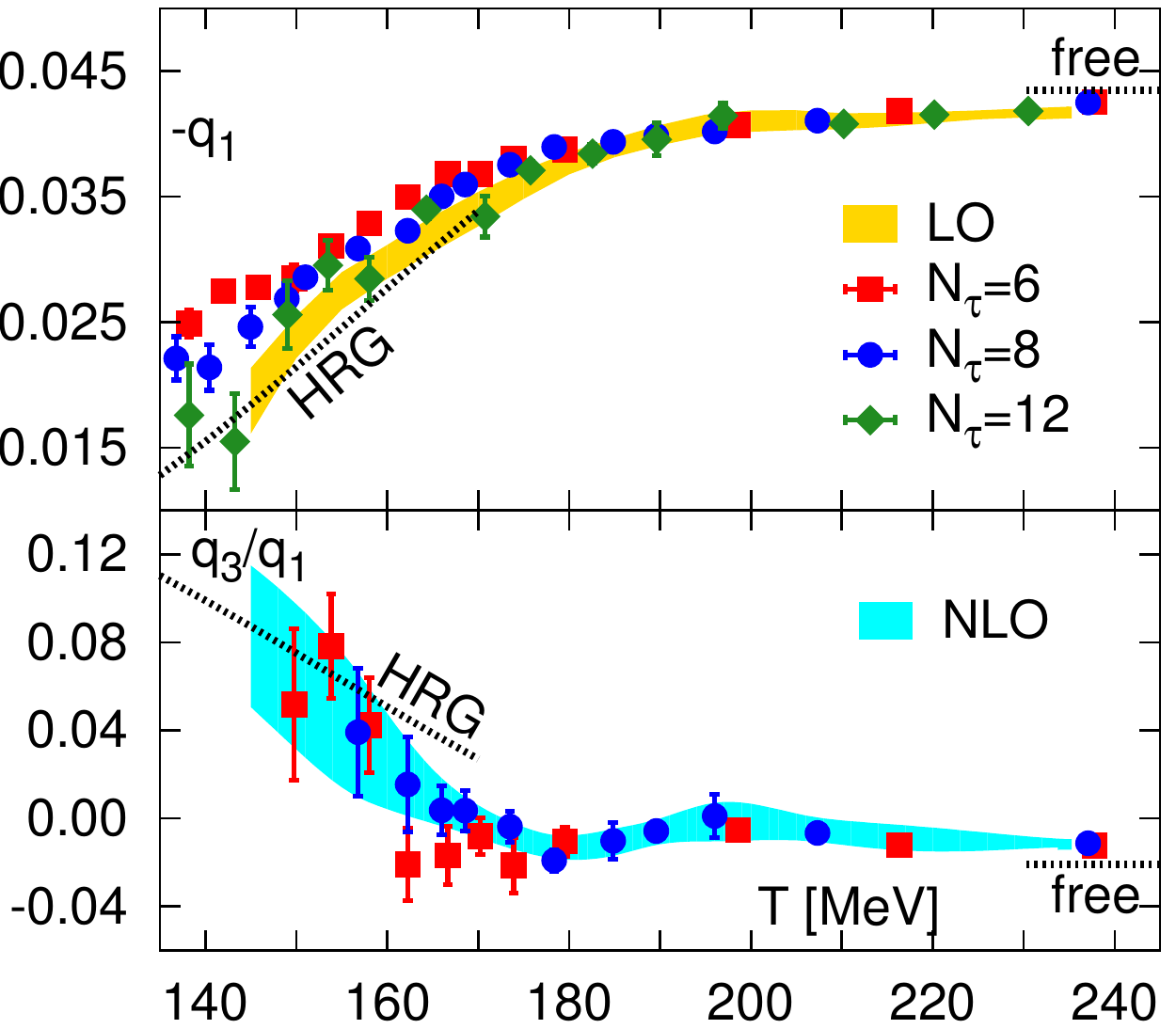} }
\subfigure[]{ \label{fig:si} \includegraphics[scale=0.32]{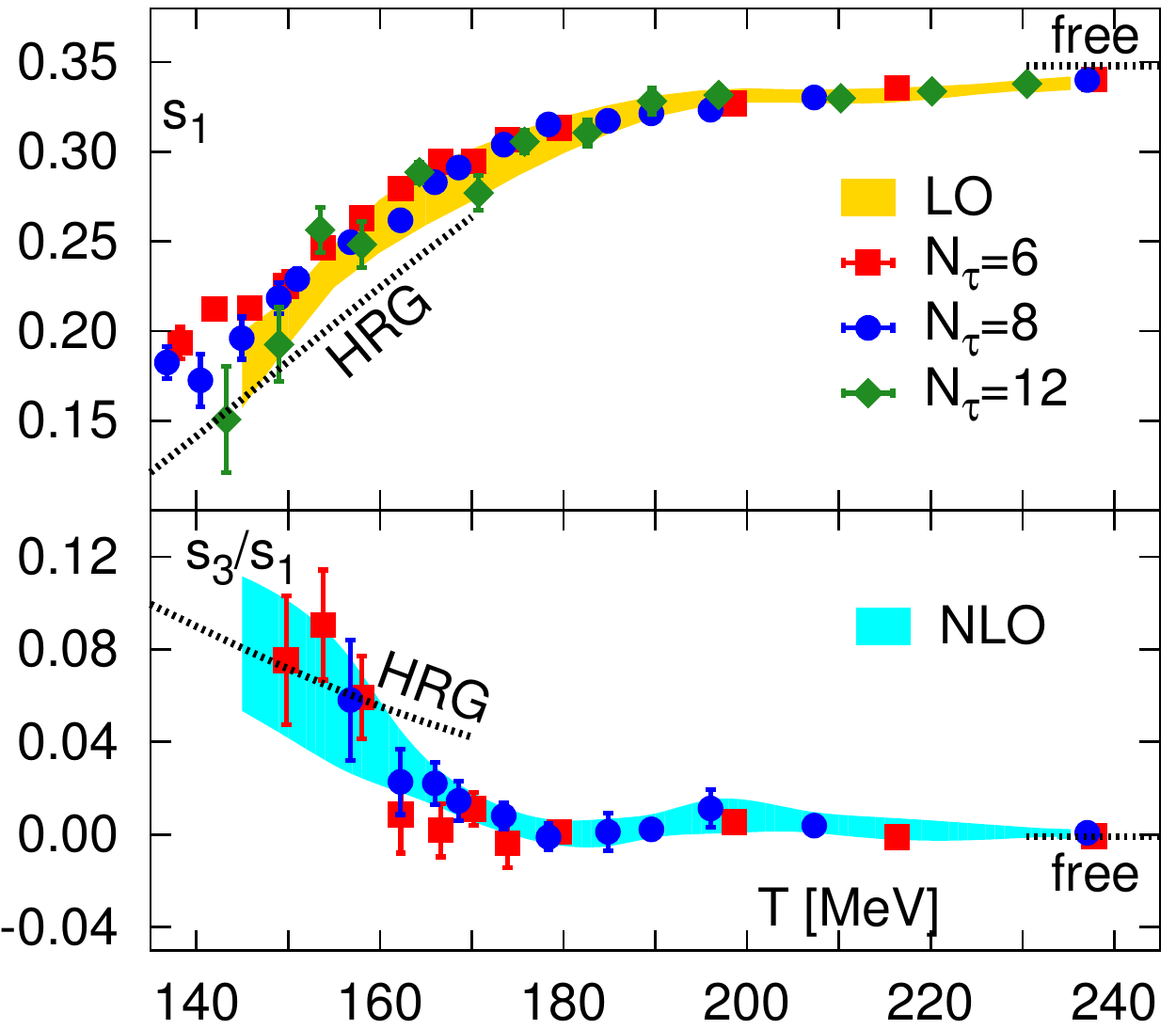} }
\subfigure[]{ \label{fig:muQS} \includegraphics[scale=0.32]{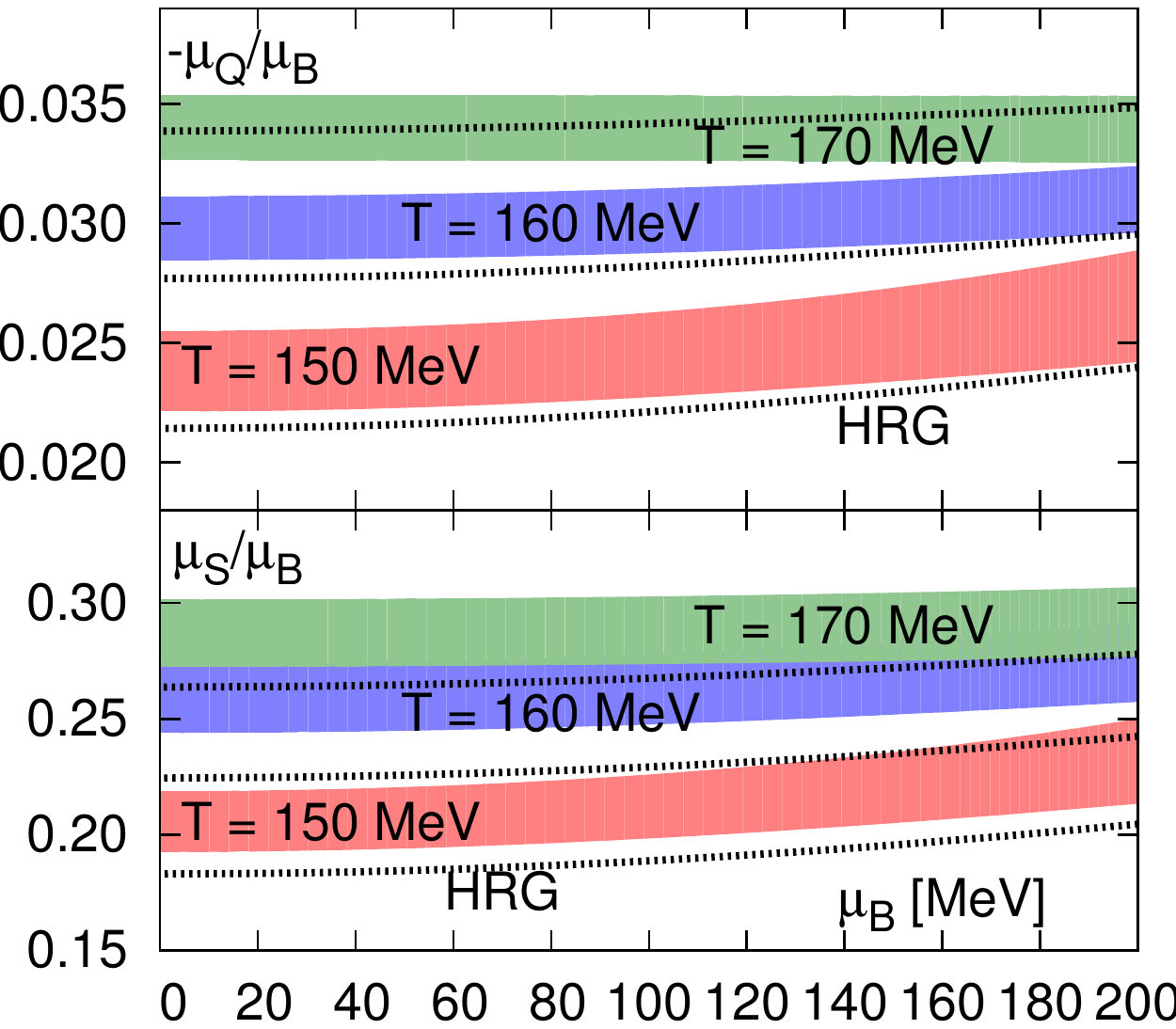} }
\caption{(a) Top: Continuum extrapolated LO in $\mu_B$ contribution for the electric
charge chemical potential as a function of temperature. Bottom: NLO contribution, in
units of the LO contribution, as a function of temperature for the electric charge
chemical potential. (b) Same as the previous panel, but for the strangeness chemical
potential. (c) Electric charge (top) and strangeness (bottom) chemical potential as a
function of $\mu_B$ for the relevant temperature range $T=150-170$ MeV.}
\efig

By using the constraints of initial strangeness neutrality and initial iso-spin
asymmetry of the colliding nuclei of HIC the electric charge and strangeness chemical
potentials, $\mu_Q$ and $\mu_S$ respectively, can be treated as dependent parameters
of $T$ and $\mu_B$. Assuming spatial homogeneity, the initial strangeness neutrality
gives $\lang n_S \rang=0$ and the initial iso-spin asymmetry of the colliding nuclei
leads to  $\lang n_Q \rang=r\lang n_B \rang$. Here, $n_X$ denotes the density of the
corresponding net conserved charge $X$ and $r=N_p/(N_p+N_n)$ is the ratio of the
total number of protons to the total number of protons and neutrons of the initially
colliding nuclei. We choose to work with $r=0.4$, a good approximation for $Au-Au$
as well as $Pb-Pb$ collisions. By making Taylor expansions of $\lang n_X \rang$ in
$(\mu_B,\mu_Q,\mu_S)$ up to $\mathcal{O}(\mu_X^3)$ and imposing the above two
constraints one can write down the $\mu_Q$ and $\mu_S$ in terms of the other two
independent parameters: $\mu_Q(T,\mu_B)=q_1(T)\mu_B+q_3(T)\mu_B^3$ and
$\mu_S(T,\mu_B)=s_1(T)\mu_B+s_3(T)\mu_B^3$.

In \fig{fig:qi} we show LQCD results for the Leading Order (LO) contribution $q_1(T)$
(top) and the Next-to-Leading Order (NLO) contribution $q_3(T)$ (bottom) to $\mu_Q$.
Similar contributions for the $\mu_S$ are shown in \fig{fig:si}. These results show
that the NLO contributions are less than 10\% and well under control for a wide range
of the baryon chemical potential $\mu_B\lesssim200$ MeV, \ie\ for RHIC energies down
to $\sqrt{s_{NN}}\gtrsim19.6$ GeV. In \fig{fig:muQS} we show $\mu_Q(T,\mu_B)$ (top)
and $\mu_S(T,\mu_B)$ (bottom) as a function of $\mu_B$ for the relevant temperature
range $T=150-170$ MeV. Comparisons with the HRG model suggest that the LQCD results
differ by 10-15\%.

\section{Freeze-out temperature and baryon chemical potential}

\bfig
\subfigure[]{ \label{fig:R31Q} \includegraphics[scale=0.32]{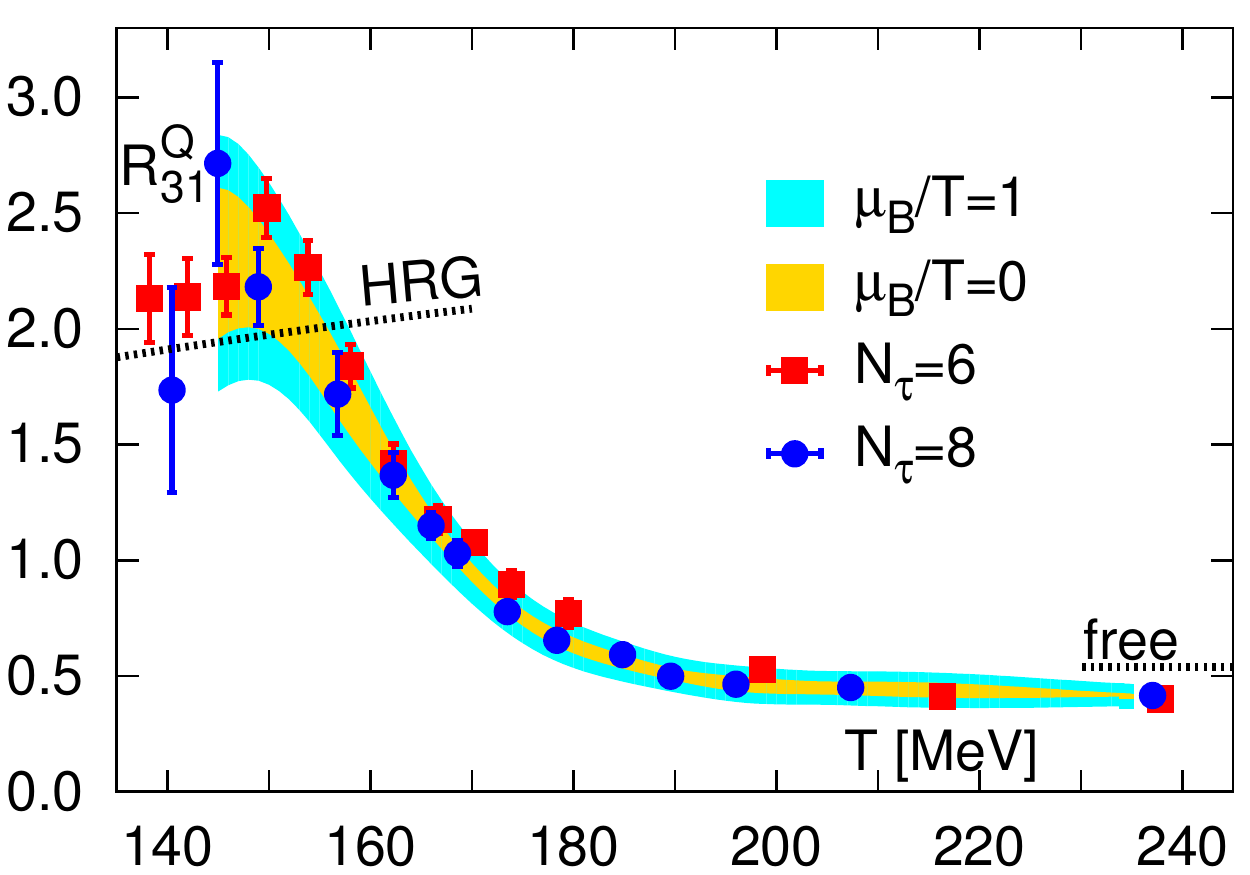} }
\subfigure[]{ \label{fig:R12Q} \includegraphics[scale=0.32]{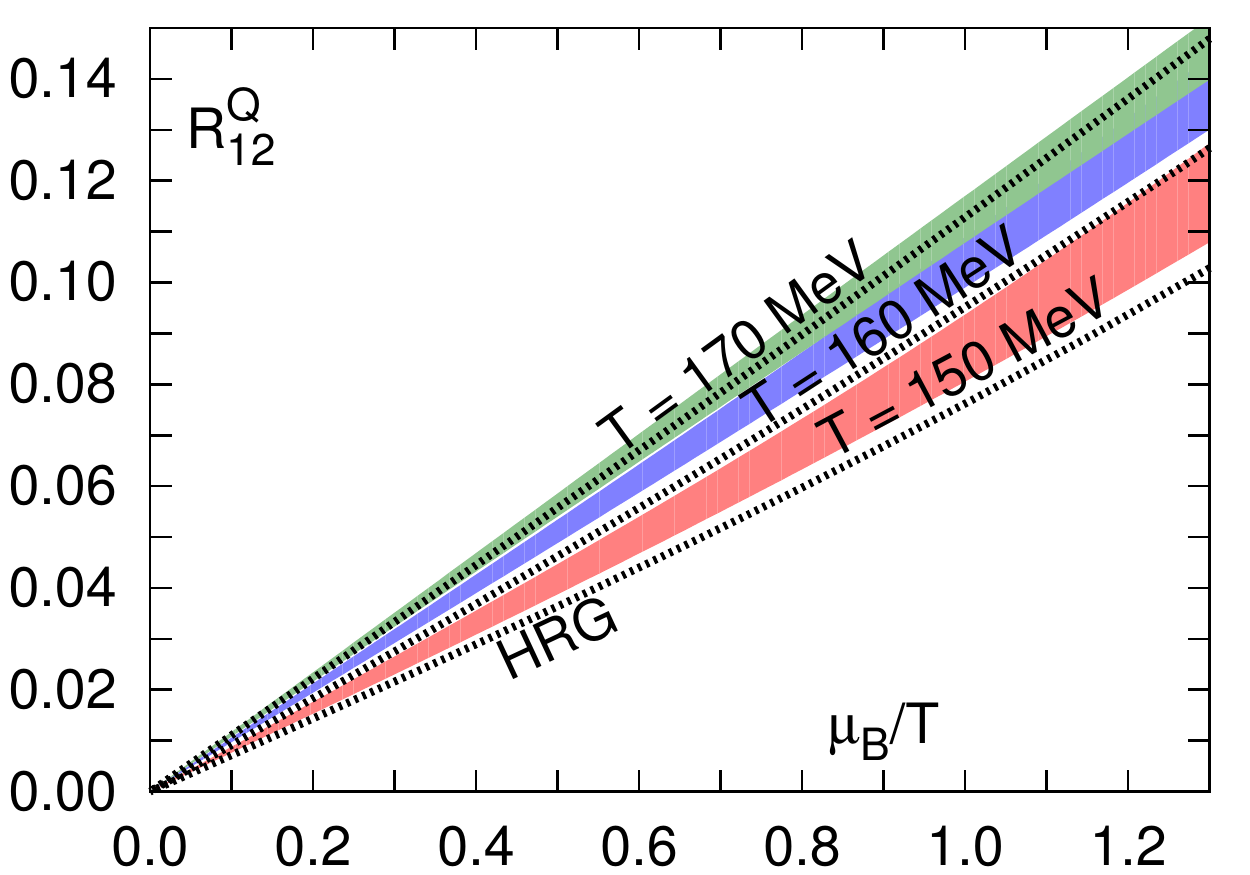} }
\subfigure[]{ \label{fig:R12QB} \includegraphics[scale=0.32]{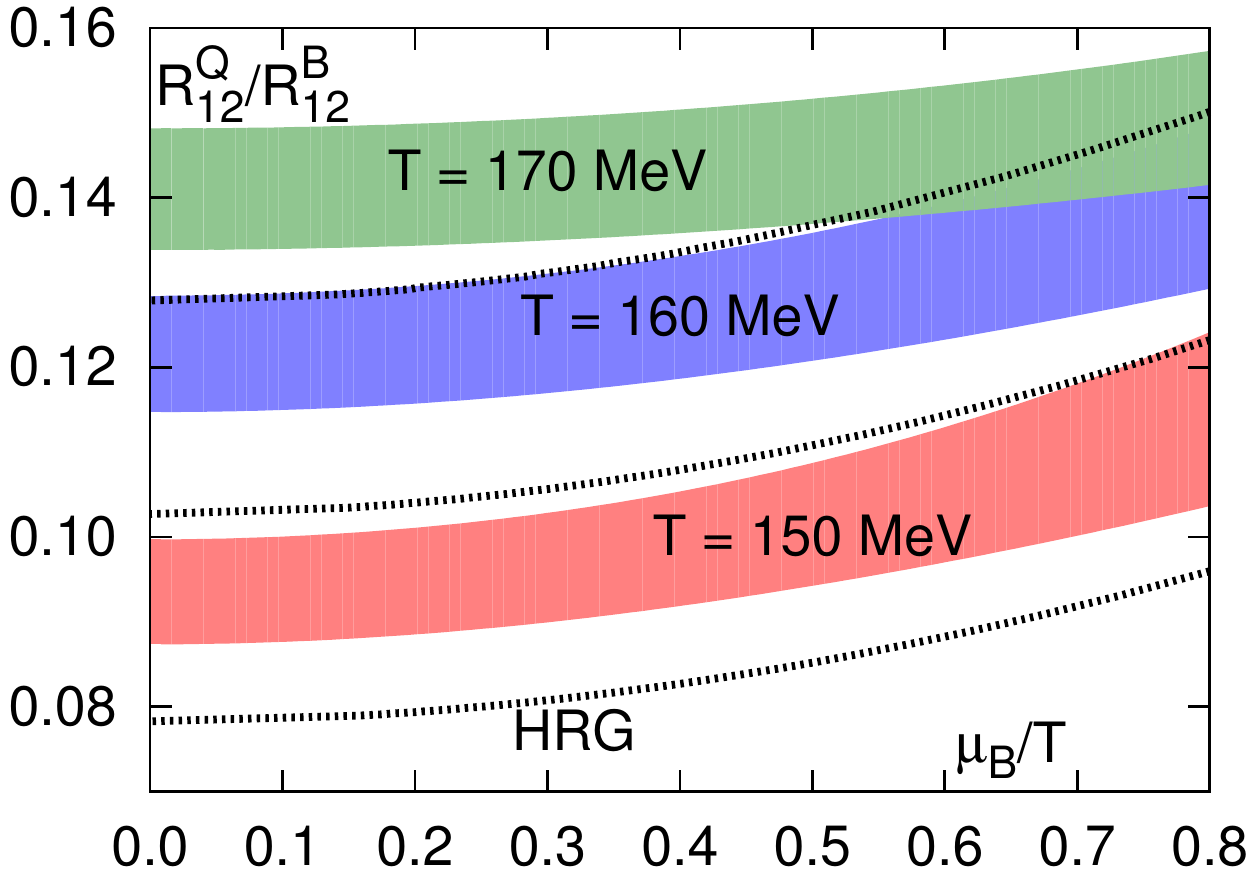} }
\caption{LQCD results for the {\sl thermometer} $R_{31}^Q$ (a) and the {\sl
baryometer} $R_{12}^Q$ (b). (c) LQCD results for the double ratio $R_{12}^Q/R_{12}^B$
as a function of $\mu_B/T$.}
\efig

As $\mu_Q$ and $\mu_S$ are known as a function of $T$ and $\mu_B$ all the cumulants
of conserved charge fluctuations can be now be expressed only in terms of $T$ and
$\mu_B$. Although in RHIC both net proton and net electric charge fluctuations are
being measured \cite{star-proton,star-charge,phenix-charge}, in LQCD only
fluctuations of conserved quantities are accessible. Since the net proton
fluctuations may not be quantitatively equal to the net baryon number fluctuations,
it is safer to work with the net electric charge fluctuations. We propose to look at
two different volume independent ratios formed out of the three lowest order
cumulants, mean ($M_Q$), variance ($\sigma_Q$) and skewness ($S_Q$), of the charge
fluctuations---
\bsubeq
\beqa
R_{31}^Q &\equiv& \frac{\chi_3^Q(T,\mu_B)}{\chi_1^Q(T,\mu_B)} =
\frac{S_Q\sigma_Q^3}{M_Q} = R_{31}^{Q,0} + R_{31}^{Q,2} \mu_B^2 + \cdots
\label{eq:R31Q} \\
R_{12}^Q &\equiv& \frac{\chi_1^Q(T,\mu_B)}{\chi_2^Q(T,\mu_B)} = \frac{M_Q}{\sigma_Q^2} 
= R_{12}^{Q,1} \mu_B + R_{12}^{Q,3} \mu_B^3 + \cdots
\label{eq:R12Q} \;.
\eeqa
\esubeq
While the cumulants of the net charge distribution $M_Q,\sigma_Q,S_Q$ are being
measured in RHIC BES \cite{star-proton,star-charge,phenix-charge}, the generalized
higher order charge susceptibilities
$\chi_n^Q(T,\mu_B)=\partial^n\ln\mathcal{Z}/\partial(\mu_B/T)^n$ can be calculated
using LQCD via Taylor expansions
$\chi_n^Q(T,\mu_B)=\sum_{k=0}\chi_{n+k}^Q(T,0)(\mu_B/T)^n/k!$. In a similar manner we
can Taylor expand the ratios $R_{31}^Q$ and $R_{12}^Q$ themselves up to NLO in
$\mu_B$ as shown in \eq{eq:R31Q} and \eq{eq:R12Q} respectively. Note that, in
the LO the ratio $R_{31}^Q$ is independent of $\mu_B$ but the LO term for the ratio
$R_{12}^Q$ is proportional to $\mu_B$. Thus, $R_{31}^Q$ can be used as the {\sl
thermometer} to determine $T^f$ and $R_{12}^Q$ can used as the {\sl baryometer} to
fix $\mu_B^f$.

In \fig{fig:R31Q} we show the LQCD results for the ratio $R_{31}^Q$. The estimated
NLO correction to this ratio is at most 10\% for $\mu_B/T\approx1$ over the whole $T$
range. The broader band in the figure depicts the range of $R_{31}^Q$ for $\mu_B/T=1$
including the NLO correction and the thinner band depicts the LO results, \ie\ for
$\mu_B=0$. The LQCD calculations for the {\sl thermometer} $R_{31}^Q$ shows a
characteristic $T$ dependence and large deviations from the HRG results within the
relevant temperature range $T=150-170$ MeV. For example, by comparing with these QCD
results an experimentally measured value of $R_{31}^2\gtrsim2$ will indicate a
freeze-out temperature $T^f\lesssim150$ MeV, $R_{31}^Q\approx1.5$ will give
$T^f\approx160$ MeV and $R_{31}^Q\lesssim1$ will mean $T^f\gtrsim170$ MeV. 

After fixing $T^f$ from the {\sl thermometer} $R_{31}^Q$ we can use the {\sl
baryometer} $R_{12}^Q$ to determine the freeze-out baryon chemical potential
$\mu_B^f$. \fig{fig:R12Q} shows LQCD results for the ratio $R_{12}^Q$ as a function
of $\mu_B$ in the relevant temperatures interval $T=150-170$ MeV using continuum
extrapolated LO results and adding contributions up to NLO in $\mu_B$. In this
temperature range the NLO corrections are well under control, less than 10\%, for
$\mu_B\lesssim200$ MeV, \ie\ for RHIC BES energies of $\sqrt{s_{NN}}\gtrsim19.6$ GeV.
By comparing the experimentally measured values of $R_{12}^Q$ with these first
principle QCD calculations one can determine $\mu_B^f$.  As a concrete example,
choosing $T^f=160$ MeV, an experimental value of $R_{12}^Q=0.01-0.02$ will give
$\mu_B^f/T^f=0.1-0.2$, $R_{12}^Q=0.03-0.04$ will suggest  $\mu_B^f/T^f=0.3-0.4$ and
$R_{12}^Q=0.05-0.08$ will indicate $\mu_B^f/T^f=0.5-0.8$.

\section{Thermodynamic consistency}

Such a determination of the freeze-out parameters for a given beam energy will help
us in understanding  whether at that beam energy higher order cumulants of conserved
charge fluctuations can be consistently described within the framework of equilibrium
thermodynamics. If the experimentally measured fluctuations are indeed described by
equilibrium thermodynamics characterized by unique values of temperature and chemical
potential then other volume independent ratios of cumulants must also have unique
values as predicted by (L)QCD calculations at the same $(T^f,\mu_B^f)$. As an
example. in \fig{fig:R12QB} we show the LQCD predictions for the double ratio
$R_{12}^Q/R_{12}^B$ as a function of $\mu_B/T$. For a pre-determined values of the
freeze-out parameters $T^f$ and $\mu_B^f$ this ratio has a unique value consistent
with equilibrium thermodynamics.  If the experimentally measured observables contain
the correct physics of equilibrium thermodynamic fluctuations then they must agree
with this (L)QCD prediction. 

\section*{Acknowledgments}

The author is supported by contract DE-AC02-98CH10886 with the U.S. Department of
Energy.


\end{document}